\documentclass{aa}
\usepackage{graphicx}
\usepackage[varg]{txfonts}
\usepackage{lscape}
\usepackage{multicol}
\usepackage{color} 
\usepackage[colorlinks, linkcolor=blue,
urlcolor=blue,citecolor=blue,bookmarks, bookmarksopen,
bookmarksnumbered,pdfauthor={Chelli et al}, breaklinks=true]{hyperref} 
\usepackage{epstopdf}
\usepackage[T1]{fontenc}
\newcommand\T{\rule{0pt}{2.6ex}}       
\newcommand\B{\rule[-1.2ex]{0pt}{0pt}} 
\begin{document}
\title{Pseudomagnitudes and Differential Surface Brightness: Application
  to the apparent diameter of stars}
\author{Chelli, Alain\inst{1}, Duvert, Gilles\inst{2,3}, Bourgès,
  Laurent\inst{2,3}, Mella, Guillaume\inst{2,3}, Lafrasse,
  Sylvain\inst{2,3}, Bonneau, Daniel\inst{1}, Chesneau, Olivier\inst{1}}
\institute{
Laboratoire Lagrange, Université Côte d'Azur, Observatoire de la Côte
d'Azur, CNRS, Blvd de l'Observatoire, F-06304 Nice, France.
  \thanks{Correspondence: {\tt Alain.Chelli@oca.eu}}
\and
Univ. Grenoble Alpes, IPAG, F-38000 Grenoble, France
\and
 CNRS, IPAG, F-38000 Grenoble, France }
\date{Received; Accepted}

\abstract{
 The diameter of a star is a major observable that serves to test the
 validity of stellar structure theories. It is also a difficult
 observable that is mostly obtained with indirect methods since the stars are
 so remote. Today only~$\sim\!600$ apparent star diameters have
 been measured by direct methods: optical interferometry and lunar
 occultations. Accurate star diameters are now required in the new
 field of exoplanet studies, since they condition the planets' sizes in
  transit observations, and recent publications illustrate a
 visible renewal of interest in this topic. 

 Our analysis is based on the modeling of the
 relationship between measured angular diameters and photometries. It
 makes use of two new reddening-free concepts: a distance indicator
 called pseudomagnitude, and a quasi-experimental observable that is independent of distance and
 specific to each star, called the differential surface brightness
 (DSB). The use of all the published measurements of
 apparent diameters that have been collected so far, and a careful modeling of the
 DSB allow us to estimate star diameters with a median statistical
 error of~$1.1\!\%$, knowing their spectral type and, in the present
 case, the VJHKs photometries. 

 We introduce two catalogs, the JMMC Measured Diameters Catalog (JMDC), containing measured star
 diameters, and the second version of the JMMC Stellar Diameter
 Catalog (JSDC), augmented to about $453\,000$ star diameters. Finally,
 we provide simple formulas and a table of coefficients to quickly
 estimate stellar angular diameters and associated errors from (V, Ks)
 magnitudes and spectral types.  }
\keywords{stars: fundamental parameters -- techniques: data analysis
  -- techniques: interferometric -- astronomical database:
  miscellaneous -- catalogs}
\titlerunning{On the apparent diameter of stars}
\authorrunning{Chelli et al.}
\maketitle
\section{Introduction}
Most of our knowledge of the Universe still comes from, either directly or
indirectly,  an analysis of starlight. Stellar physics is used,
implicitly or explicitly, in every astrophysical research field, and
must be continually tested against observations. One of the most
fundamental and dimensioning parameters of the physics of a star is its
size, measured through its distance and its apparent angular size.
The original interest in measuring stellar angular sizes was to get
precise estimates of the stellar effective temperature, $T_{\mathrm{eff}}$,
which is pivotal to, for example, determining ages and
metallicities. But there is renewed interest, such as the precise
stellar radii now needed to measure transiting exoplanets' sizes and
densities, which can eventually fall into the host star's so-called
habitable zone \citep{2014MNRAS.438.2413V}.

Primary distance measurements are the realm of the space missions
Hipparcos~\citep{ESAHIP} and GAIA~\citep{GAIA}. The most
straightforward method to measure the angular sizes of stars is
interferometry~\citep{Michelson,HanburyBrownAndTwiss,Labeyrie}, and in
a less measure, more historically, lunar occultations
\citep{Cousins50}. Modern interferometric techniques -- combining
adaptive optics, mononode fibers, cophasing, fast and low noise NIR
cameras -- are today able to measure stellar diameters with a
precision better than~1\% \citep{2013MNRAS.434..437C,
  2012ApJ...746..101B,2012ApJ...757..112B}. However the ultimate
precision is often dominated by the calibration process. To reach 
very high precision, it is mandatory to know and to observe with the
state of the art procedures, together with the object of interest, a star called
calibrator. In other words, a single star, as close as possible to the
object, with similar magnitude, with a known angular diameter and
associated error, that is used to measure the transmission of the
observing system. This underlines the importance of angular diameter
predictions for calibration purposes. The angular diameter predictions are
based on two kind of methods:
the first  are based on a
polynomial fit of measured angular diameters as a function of 
colors, from ~\cite{1969MNRAS.144..297W} 
to \cite{2013ApJ...771...40B,2014AJ....147...47B}; the second,
like the Infrared Flux Method \citep{1979MNRAS.188..847B,
  2010A&A...512A..54C},  are
based on a subtle mix of experimental data and modeling.
Other methods, like Asteroseismology
\citep{1994ARA&A..32...37B,2013ARA&A..51..353C}, provide linear
diameters that 
must be converted to angular diameters using a distance estimate.

Our initial motivation for this work was to optimize the
calibrator-finding utility SearchCal\footnote{
  \url{http://www.jmmc.fr/searchcal}}~\citep{Bonneau1,Bonneau2}, an
angular diameter estimator of the first kind described above, with a
rigorous treatment of error propagation. The aim was to produce a more
robust version of the catalog of~$\sim40\,000$
star diameters, the JMMC Stellar Diameter
 Catalog (JSDC)~\citep{JSDC}. This
led us to reconsider the general problem of deriving the polynomial
coefficients that are used in such estimators, which usually come from star magnitudes that
need to be de-reddened before use. We solved the visual extinction
problem by introducing two new concepts: a distance indicator called
pseudomagnitude, and the differential surface brightness
(hereafter DSB), both reddening-free.  The DSB is an observable specific
to each star, independent of distance, which 
depends only on measured quantities: the stellar diameter and the
observed magnitudes.

Our approach to predict stellar diameters, definitely experimental,
consists of a polynomial fit of the DSB as a function of the spectral
type number for a database of stars with known diameters. It
allows us to by-pass the knowledge of visual extinction and that of  intrinsic colors. The
polynomial coefficients are then applicable to any star with a known
spectral type and magnitudes to provide an apparent diameter value.
As an illustration of our method, we compiled a catalog of~$\sim600$
measured star diameters and photometries that we used to compute the
DSB polynomial fit. Using these polynomials and all the stars in the
ASCC catalog \citep{ASCC} that have an associated spectral type, we are able to give the
apparent diameter of~$\sim453\,000$
stars with a median diameter precision of $1.1\%$,
as well as possible astrophysical biases up to $2\%$ (due to luminosity
classes, DSB fine structures, metallicity, and so on), that is in agreement
with the limiting precisions discussed in ~\cite{2014MNRAS.439.2060C}.

Section~\ref{sec:formalism} introduces the new formalism,  
especially the concepts of pseudomagnitudes and differential surface
brightness. Section~\ref{sec:algo} describes the least squares fit
approach. Section~\ref{sec:database} describes how we built the database of measured
diameters from the literature, discusses their validity, and how we tried to avoid any systematics.
The results are presented and discussed in Section~\ref{sec:results}.   
\section{Pseudomagnitude and differential surface brightness}\label{sec:formalism}
In this section, we introduce the concept of pseudomagnitude and a new
experimental observable: the differential surface brightness (DSB).  
Our starting point is the expression of the surface brightness
$\mathrm{S}_i$ of a star \citep{1969MNRAS.144..297W}, see Equation 9 of
\citep{1976MNRAS.174..489B}: 
\begin{equation}
\mathrm{S}_i = 5\log(\theta)+m_i^0
,\end{equation}
where $\theta$ is the angular diameter of the star and $m_i^0$ the
unreddened magnitude in the photometric band $i$. $m_i^0$ can be written as:
\begin{equation} 
m_i^0 = m_i-c_iA_v
,\end{equation}
where $m_i$ is the observed magnitude, $c_i$ is the ratio between the
extinction coefficients $R_i$ and $R_v$ in the $i$ and visible bands,
respectively. Given two photometric bands $i$ and $j$, the
interstellar extinction can be writtten as
\begin{equation} 
 A_v = \frac{m_i-m_j}{c_i-c_j} - \frac{m^0_i-m^0_j}{c_i-c_j} 
,\end{equation}
Combining the three previous equations, the surface brightness may be
expressed as a function of the angular diameter, the measured
magnitudes, and the intrinsic color of the star, $C_{ij}=
m^0_i-m^0_j$. That is to say
\begin{equation} 
\mathrm{S}_i= 5 \times \Bigg ( \log(\theta) +
0.2 \times \frac{c_i\,m_j-c_j\,m_i}{c_i-c_j}\Bigg ) +\frac{c_i}{c_i-c_j}\,C_{ij}
\label{surface brightness}
.\end{equation}
At this point, it is useful to introduce a new stellar observable that
we call pseudomagnitude, defined by
\begin{equation}
 {pm}_{ij}=\frac{c_i\,m_j-c_j\,m_i}{c_i-c_j}
.\end{equation}
The pseudomagnitudes have remarkable properties and
applications that will be discussed in a forthcoming paper. Basically,
they are reddening free distance indicators: 
\begin{equation}
{pm}_{ij}=\mathrm{PM}_{ij}+d_M
,\end{equation}
where $\mathrm{PM}_{ij}$ is the absolute pseudomagnitude, i.e., the
pseudomagnitude at $10$\,pc distance, $d_M=5\log(d)-5$ is the distance modulus and $d$
the distance in parsecs. We note that if one of the coefficients $c_i$
or $c_j$ tends to zero, then the pseudomagnitude tends to the magnitude $m_i$
or $m_j$. Hence, the absolute pseudomagnitude is, in some way, related to
the stellar luminosity.

The surface brightness may simply be rewritten as follows:
\begin{equation} 
\mathrm{S}_i= 5\times{\mathrm{DSB}}_{ij} +\frac{c_i}{c_i-c_j}\,C_{ij}
\label{surf brightness}
,\end{equation}
where ${\mathrm{DSB}}_{ij}$ is the DSB between the
photometric bands $i$ and $j$, defined as
\begin{equation}
{\mathrm{DSB}}_{ij}=\log(\theta) + 0.2\,{pm}_{ij} 
\label{eq_diam}
.\end{equation}
%
%
${\mathrm{DSB}}_{ij}$ is a self-calibrated observable specific to each star. It is 
reddening-free, independent of the distance, and can be measured via photometry and
interferometry. The only a~priori is the diameter limb-darkening
correction, which induces a diameter error that is generally less than~1\%,
smaller than the errors managed in this work.

The knowledge of the DSB, as a function of the
spectral type, the luminosity class, 
the metallicity, and so on, is sufficient to predict the angular
diameter of a star, given its observed magnitudes. Unfortunately, we
do not possess this level of   detailed information because of the poor spectral type-sampling owing to the small number of useable measured
diameters ($\sim 600$) available today. To compensate our lack of knowledge, we fit
the DSB as a function of the spectral type number
$n_s$ (varying from $0$ to $69$ for spectral types between
O$0$ to M$9$) with simple polynomial laws. Within the present
framework, if the determination of spectral type
is robust to reddening (e.g., is not derived from colors), then the
intrinsic color, replaced by the spectral type number, is no longer
a variable of the diameter prediction problem and our approach is fully
reddening-free.
\section{Algorithmic approach}\label{sec:algo}
This is a two-step process: 1) polynomial
estimate based on the DSBs derived from database values, 2) diameter calculations from polynoms
and pseudomagnitudes.
\subsection{Polynomial}
\label{sec:polynomial}
A data point consists in a limb-darkened diameter $\theta$ and $N_B$
photometric bands, which provides $N_B-1$ linearly independent (but
statistically dependent) equations. We can build many
sets of $N_B-1$ linearly independent equations, but each set
 provides the same polynomial solution. We then have  to
characterize, $N_B-1$ polynomials of degree $m$, each associated with a 
pair of photometric bands. This corresponds to $(m+1) \times
(N_B-1)$ unknowns, that we evaluate simultaneously via a simple linear
least squares fit (see Appendix~\ref{app:linearsqfit}). 

Wherever possible, we used the mean interstellar
extinction coefficients of \cite{1999PASP..111...63F} that give:
$(c_V,c_J,c_H,c_{Ks})=(1.0,0.28,0.17,0.12)$. Strictly speaking,
the coefficients $c_i$ depend on the spectral type \citep{2004AJ....128.2144M}. In
principle, one can consider variable coefficients, but for the present
analysis we restrict ourselves to constant values, since the influence
of these variations, with respect to diameter calculation, is smaller
than the errors and the biases reported here. 

Also, it may be necessary to reject anomalous data
points.  For example, in the production of the star diameter catalog
described in Sect.~\ref{sec:results}, we use the database described in
Sect.~\ref{sec:database}. From the found polynomial solution, for each entry of our database we can
estimate  $N_B-1$
single diameters and a mean diameter. It sometimes occurs that one or various
single diameters deviate from the mean diameter by more than~$5$
times the error on the difference. This may be due to observational
biases in the diameter measurements, or that the star is not
single, etc. In this case, we excluded the entry from the 
database that was used. Table~\ref{starlist} lists all the stars and references
retained for fitting the polynomials, as discussed in
Sect.~\ref{sec:results}. 
\subsection{Diameters}
A polynomial, together with two magnitudes (pseudomagnitude), provides an
estimate of the diameter. Several pairs of magnitudes gives several
estimates of the same diameter. These are not statistically
independent and we show in Appendix \ref{app:diamcalc} how to
rigorously combine them in a single diameter value and its associated
error. Beyond the error improvement, which is generally modest, the
main interest of using more than two photometric bands is to produce
various diameter estimates that can be recombined and, above all, compared through a
quality factor, see Section~\ref{sec:results}.
\section{The database of measured angular diameters}\label{sec:database}
\subsection{Building the database.}
The empirical approach used in this paper relies on the knowledge of a
statistically significant number of accurately measured angular
diameters, ideally obtained with different techniques, such as avoiding
as much as possible any technique-related bias. The angular diameters must be
prime results, i.e., not the result of a modeling of the
observations. Ideally, they should  be equally distributed, both in
space and in spectral type. 

Several star diameter compilations exist that contain a fair
amount of published angular diameters values. The CADARS
\citep{CADARS} has entries for~$6888$ stars and is complete up
to~$1997$. CHARM2 \citep{CHARM2} lists~$8231$ measurements of~$3243$
stars, up to~$2005$. However these catalogs mix results from very
direct methods, 
such as intensity interferometry with indirect methods, or
spectrophotometric estimates of various kind (always including some
model of the star), or linear diameters from eclipsing binaries
($1600$ entries in CADARS), which need some modelling of the two
stars, as well as a good estimate of the distance to be converted into
an angular diameter. 

Another difficulty is that, the published angular diameters have been
obtained at various wavelengths and may include, or not, a compensation
for the limb-darkening effect. As a result, we initiated a new
compilation of measured stellar diameters that would suit our needs.
This database, called JMDC\footnote{\texttt{JMMC Measured stellar
    Diameters Catalog}, available at \url{http://www.jmmc.fr/jmdc}},
only uses  direct methods, merges multiwavelength measurements into one
value of limb-darkened diameter (LDD), and aims to be complete up to
the most recent publications by being updated on a regular basis
through a peer-reviewed submission process.

The JMDC used in this paper gathers~$1072$ apparent diameter values 
that have been published since the first experiments by Michelson. Prior
to~$1997$, our bibliography relies only on the reference list of 
\cite{CADARS}. After this date we used NASA's ADS hosted at CDS.
We retained only the measurements obtained from visible/IR
interferometry, intensity interferometry and lunar occultation
 in the database. We always retrieved the values in the original
text\footnote{with the exception of a few very old references in
  CADARS that were not easily available at our location.} and used
SIMBAD to properly and uniquely identify the stars.

The three techniques retained share the same method of converting the
measurements (squared visibilities for optical interferometry,
correlation of photon-counts for intensity interferometry, fast
photometry for lunar occultations) into an angular diameter: fitting a
geometrical function into the values, in many cases a uniform disk,
which provides a uniform disk diameter (UDD) value. This UDD is
wavelength-dependent owing to the limb-darkening effect of the upper
layers of a star's photosphere, and JMDC retains the wavelength or
photometric band at which the observation was made.

To measure a star's apparent diameter consistently, i.e., with the same
meaning as our Sun's well-resolved apparent diameter, it was necessary
for the authors of these measurements to take into account the star's
limb-darkening, for which only theoretical estimates exist as yet. They
chose one of the various limb-darkening parameters available in the
literature (see \cite{CLARET2000} for a discussion on the classical
limb-darkening functions used), either by multiplying the UDD by a
coefficient function of the wavelength and the star's adopted
effective temperature, or directly fitting a limb-darkened disk model
in the data. Of course this adds some amount of theoretical bias in
the published measurements, which however diminishes as the wavelength
increases. An additional difficulty for the lunar occultations is that
the result depends on the exact geometry of the occulting portion of
the lunar limb, which can,  more or less, be correctly estimated.

To deal with the limb-darkening problem as efficiently as possible, in the
publications where reported diameters are measured in several
optical/IR bands, we retained the measurement with the best accuracy
and favored the measurement at the longest wavelength to minimize the
effect of limb-darkening correction. Furthermore, we further used the published UDD
measurement, or retrieved the original, unpublished UDD measurement
from the LDD value and the limb-darkening coefficient used by the
authors, and uniformly converted  these UDD values into limb-darkened
angular diameters using the most recent correction factors
published \citep{2013A&A...554A..98N,2013A&A...556A..86N}, when
possible. This, in our opinion, works towards minimizing the biases of
the database, which will be confirmed afterwards by the statistical
analysis of our results.

We did not keep any other information from the original references. Instead,
we  retrieved all the ancillary
information such as photometries, parallaxes, spectral types etc, at
once by using our \texttt{GetStar} service, a specially crafted
version of our \texttt{SearchCal}
server\footnote{\url{http://www.jmmc.fr/getstar}} that fetches all
relevant information from a dozen  CDS-based catalogs (VizieR,
Simbad for object and spectral types). This ensures that there is no
difference in the origin, thus no added bias, between the database we
use to derive our polynomials (see below) and those that will be used
in the reverse process, for any object known by Simbad at CDS.
\subsection{Database properties}
As of today, the database retains~$1072$ different measurements
on~$627$ stars from~$169$ publications. Of them, $204$~stars have multiple
entries, from~$2$ ($205$ stars) 
to~$18$ ($\alpha$\,Tau). In addition, $1041$~measurements have an UDD value,
$565$~reported an LDD value. After an eventual LDD to
UDD conversion  
and use of the above-mentioned conversion factors (for compatible
spectral types), we are left with~$853$ entries with useable LDD
measurements. 

With regard to the techniques, data is issued from long-baseline
optical/IR interferometry ($68\%$), then lunar occultations ($26\%$),
and finally intensity interferometry ($6\%$). Of the
measurements retained,  $36\%$ are in the K band (around~$2200$\,nm), the rest
being equally distributed between the B,V, R, I, and H bands. 
\subsection{Database final filtering}\label{sec:databasefiltering}
For the purposes  of this work (single stars of classical spectral types),
we remove the known multiples or strongly variable stars (cepheids,
miras, close doubles\footnote{Of separation less than one arc
  second.}, spectroscopic binaries, elliptical variables etc) to
obtain a set of (apparently) standard stars. In addition, we  consider only
stars with the following complete information: VJHKs magnitudes and
errors\footnote{Our GetStar service is used to retrieve Johnson V
  magnitudes from the ASCC \citep{ASCC} and Hipparcos \citep{ESAHIP}
  catalogs, and JHKs from 2MASS \citep{2MASS}. GetStar makes a careful
  cross-match on the stellar properties reported in these catalogs,
  comparing them with Simbad values, and takes into account proper
  motions in its cross-matching. }; SIMBAD spectral types within half
a subclass precision; LDD 
diameters and their errors. Finally, keeping only measurements with an
S/N above five for LDDs, we are left with 573 measurements of 404
distinct stars.

The spectral types of the  573 selected measurements range from O5 to M7 in
five luminosity classes: $124$ dwarves (V), $42$ subgiants (IV), $297$
giants (III), $38$ subsupergiants (II), $56$ supergiants (I), and
$16$ without luminosity class (in the SIMBAD database). The diameter
measurements range over more than two decades, from~$0.23$ to~$44$ mas.
\section{Results and discussion}\label{sec:results}
To derive the polynomial coefficients and their errors, we used 
the 573 selected measurements of our database (see Section~\ref{sec:databasefiltering}), and
simultaneously fitted  the DSB of the photometric
pairs (V,J), (V,H), and (V,Ks).

The quality of the polynomial
fit is evaluated by the associated chi-square, $\chi_p^2$ (see
Appendix~\ref{app:linearsqfit}), and that of
the reconstructed diameters by the mean diameter chi-square
${\langle}\chi_\theta^2{\rangle}$, (see Appendix~\ref{app:diamcalc}). 
\subsection{Influence of luminosity class}
To test the influence of the luminosity class (LC), we first computed  three sets of
polynomials of degree 6 separately, from: 1) LC I, II and III ($\chi_p^2=0.6,
{\langle}\chi_\theta^2{\rangle}=0.6$); 2) LC IV and V 
($\chi_p^2=0.8,  {\langle}\chi_\theta^2{\rangle}=0.7$); and 3) all, including unknown
LC ($\chi_p^2=0.7,  {\langle}\chi_\theta^2{\rangle}=0.7$). Then in
each case, we computed the angular 
diameters of all stars of the
Hipparcos~\citep{ESAHIP} catalog  with known spectral types and 
magnitudes ($93\,142$ stars). Within an rms bias of $2\%$, we found no  
difference between cases 1) and 3), and cases 2)  
and 3). As a consequence, we decided to use all the selected
measurements of our database, irrespective of their
luminosity class (or absence thereof), to derive a single set of polynomials that is
applicable to all stars. 
   \begin{figure}
   \centering
   \includegraphics[angle=90,width=8cm]{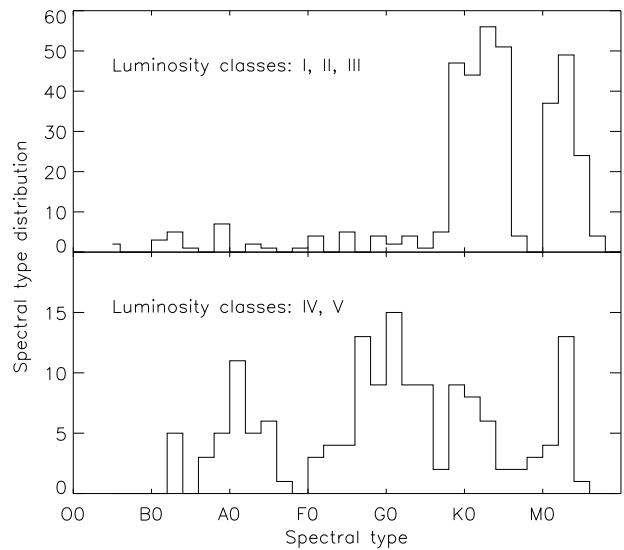}
      \caption{Histogram of retained stellar diameter measurements for
        the DSB polynomial fit. Top: Note the overabundance of
        luminosity classes I, II, and III (363 entries), around K and M spectral
        types and the poor sampling for earlier ones; Bottom: Spectral
        distribution for luminosity classes IV and V (150 entries).   
              }
         \label{fig:histo_lc}
   \end{figure}
   \begin{figure}
   \centering
   \includegraphics[angle=90,width=8cm]{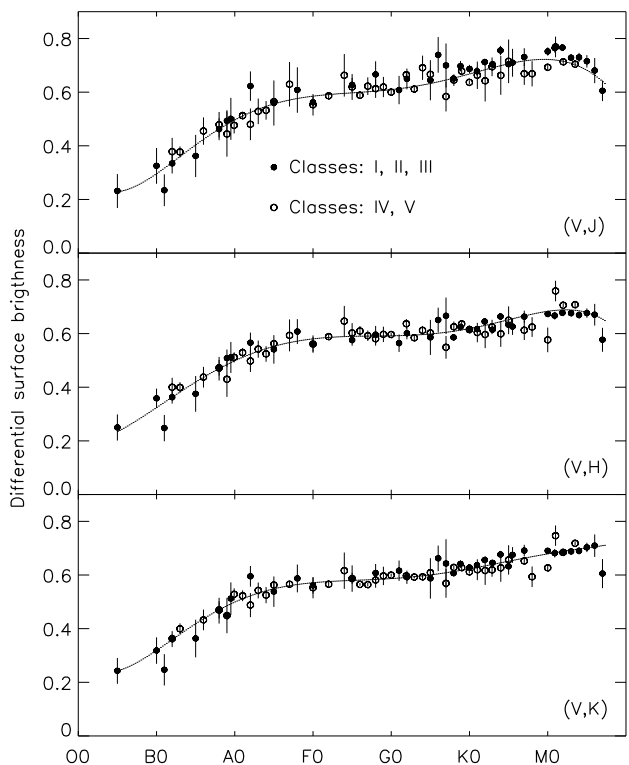}
      \caption{DSB values in
the three selected photometric pairs, together with polynomial fits of
degree 6 (dotted lines), as a function
of the spectral type. The
entries were binned separately, with an interval of one subspectral type, 
for LC I, II, and III (filled circles) and LC IV and V (open circles),
to understand their similarity. 
              }
         \label{fig:dsb_vs_sptype1}
   \end{figure}
\subsection{Selected database measurements and their DSB}
From the initial $573$ entries, $526$ ($92\%$) filled the fitting condition
(see end of Section~\ref{sec:polynomial}) and were retained for  
polynomial calculations ($363$ entries with LC~I, II, and III, $150$ with LC
IV and V, and $13$ with unknown LC). The spectral type distribution for
luminosities classes I, II, and III is shown in
Figure~\ref{fig:histo_lc}(top). There is an overabundance of
measurements around K and 
M spectral types and a poor sampling for earlier ones. Instead, for
luminosity classes IV and V (Figure~\ref{fig:histo_lc}, bottom) the
spectral types are quite well distributed between B and M types. Also
note that 100\% (31 entries) of interferometry intensity data, 92\%
(407 entries) of classical interferometry data and 87\% (88 entries) of
lunar occultation data, have been retained. 

\begin{figure}
   \centering
   \includegraphics[angle=90,width=8cm]{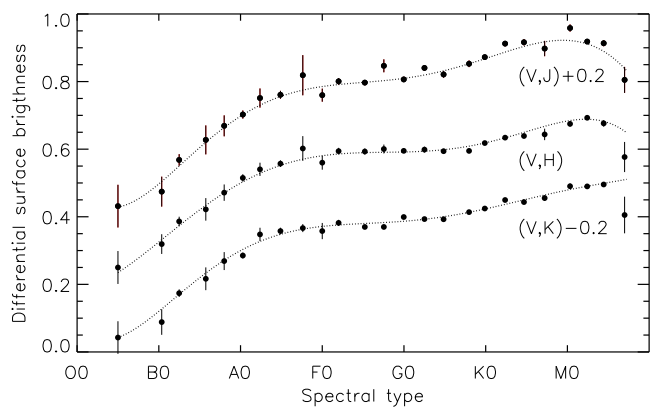}
      \caption{Differential surface brightness for (from top to bottom) the (V, J),
        (V, H) and (V, Ks),
        photometric pairs versus spectral type. The (V, J) and (V, Ks)
        curves have been shifted by $\pm 0.2$ for clarity. The
entries were binned, with an interval of 2.5 subspectral type. The
dotted lines represent the best fit with polynomials of degree 6. Note
that the spectral metric is not smooth (see text).
              }
         \label{fig:dsb_vs_sptype2}
   \end{figure}

Figure~\ref{fig:dsb_vs_sptype1} shows the DSB in
the three selected photometric pairs as a function of the spectral
type, together with the polynomial fit. The entries were binned, with
an interval of one subspectral type, separately for LC I, II and III (filled circles) and LC
IV and V (open circles), to understand their
similarity. Figure~\ref{fig:dsb_vs_sptype2} shows the same DSB with
all entries binned with an interval of 2.5 subspectral type. 
It is clear that the spectral metric is not smooth, especially for the
(V,J) pair. It 
presents a period of about one spectral type plus some possible
fine structures. In fact no polynomial of reasonable degree is able to
reproduce the exact structure of the DSB and we must always
keep in mind the $2\%$ rms bias on the diameter.
\subsection{Comparison of diameter predictions with recent measurements}
\begin{figure}
   \centering
   \includegraphics[angle=90,width=8cm]{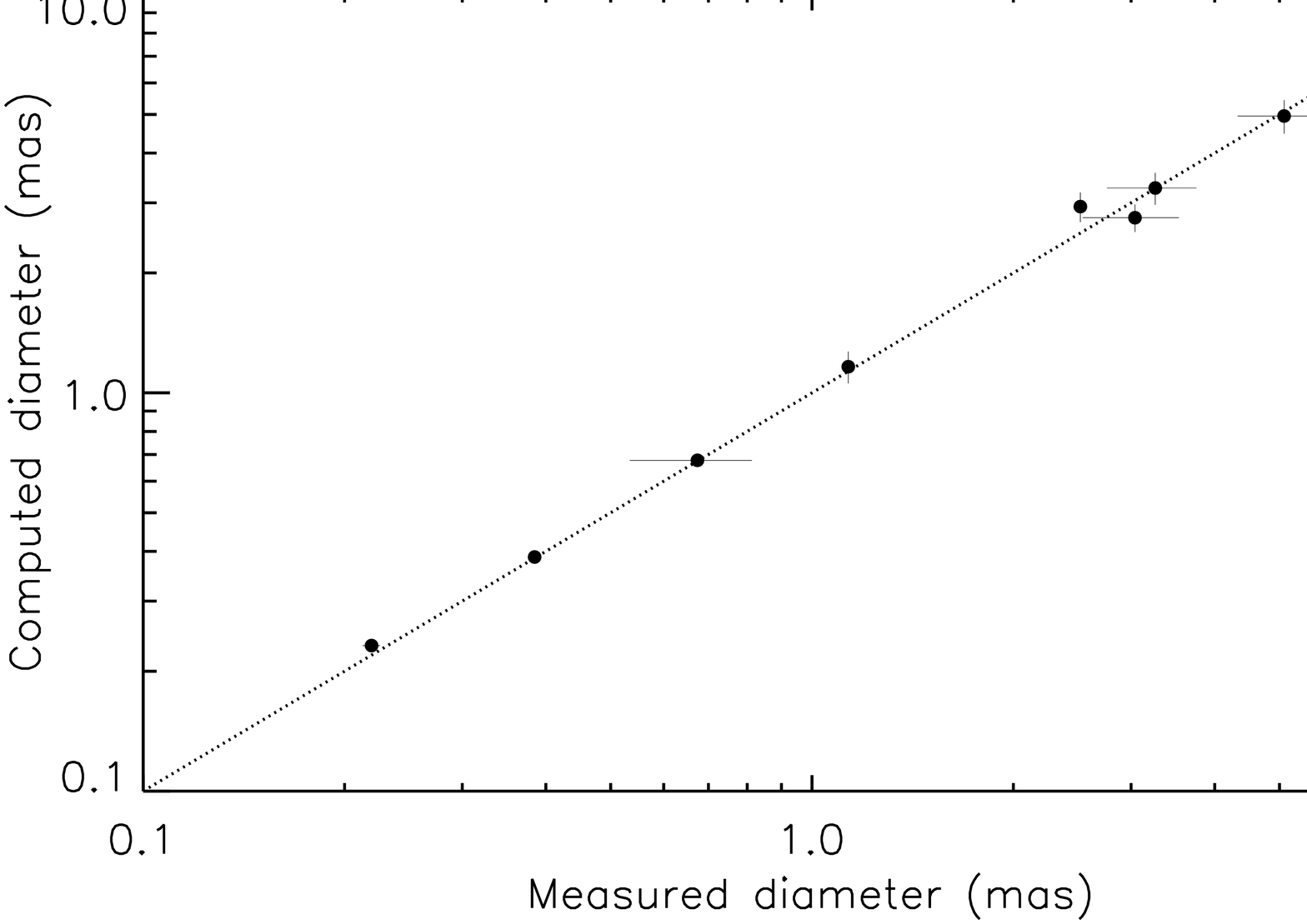}
      \caption{Computed diameters as a function of
measured diameters, for eight stars recently observed with
interferometry and not used for the polynomial fit. Stars with increasing diameter, HD
numbers: 209458 (G0V), 189733 
(K1.5V), 69830 (G8), 185351 (G9III), 190658 (M2.5III),
183589 (K5I), 95687 (M3I),  97671 (M3I), 
\citep{2014ApJ...794...15J,2014A&A...564A...1B,2015ApJ...800..115T,2015MNRAS.447..846B,2015A&A...575A..50A}. The  
agreement is excellent, with a mean-squared 
difference between measured and computed diameters expressed in
noise units of 0.5. The straight dotted line corresponds to x=y.
              }
         \label{fig:comp_vs_meas_diam}
   \end{figure}

To test the diameter predictions of our model,
we computed the diameter of eight stars, which had  recently been measured by
interferometry and not used for the polynomial
fit. Figure~\ref{fig:comp_vs_meas_diam} shows the computed diameters as a function of the
measured ones. The agreement is excellent, with a mean-squared
difference between measured and computed diameters that is expressed in
noise units of about 0.5. 
\subsection{The JMMC stellar diameter catalogue}
We used the present formalism to compute the angular diameters of
stars in the Tycho2 catalog\footnote{See the JSDC catalog at
  \url{http://www.jmmc.fr/jsdc}. The Tycho2 catalog is used here as a
  list of star positions and photometries are retrieved through
  GetStar, as described in
  Sect.~\ref{sec:databasefiltering}. }~\citep{TYCHO2} with known
spectral type. The 
resulting median diameter error is $1.1\%$.
About $453\,000$ stars have an associated internal diameter
$\chi_\theta^2$ less than~5, and $393\,000$ less than~2.
\subsection{Simplified formula}
For a star which could be absent in the JMMC catalog and for which two
tycho-like magnitudes and a spectral type are known, it is easy to
derive a diameter estimate using the formula below.

Knowing the DSB polynomial values $p$ and their errors $\sigma_p$, the
stellar diameter (mas) and its relative error may quickly be computed using the
following formulas:  
\begin{equation}
\log(\theta)=-\,0.2\times\frac{c_V\,m_X-c_X\,m_V}{c_V-c_X}+p
\label{logtheta}
\end{equation}
and
\begin{equation}
\frac{\sigma(\theta)}{\theta }=\ln (10)\,\sqrt{
0.04\times\frac{c_V^2\sigma^2(m_X)+c_X^2\sigma^2(m_V)}{(c_X-c_V)^2}+\sigma_p^2}
\label{error_theta}
,\end{equation}
with X=J, H or Ks. Table 1 provides the pair ($p$, $\sigma_p$) for
the (V,Ks) photometries and stars of spectral type O5 to M6.

It is also possible to estimate angular diameters without a 
precise knowledge of the spectral type, but with a
degraded precision. For a star whose spectral type is known within
some range, it suffices to replace in Eq.~\ref{logtheta} and
Eq.~\ref{error_theta}, the pair ($p$, $\sigma_p$) by the pair
($a, \sigma_a$), where $a$ is the mean value of $p$, and $\sigma_a$ its dispersion over
the spectral range. If the spectral
type is in the range O5--M6, then for any X, ($a, \sigma_a$)=(0.56, 0.12),
for the range A0--M6, ($a, \sigma_a$)=(0.62, 0.05), which respectively provides $28\%$ and
$12\%$ diameter error. 
\section{Conclusion}
Our approach to predict stellar angular diameters is based on the modeling of
the relationship between angular diameters and photometries.
We developed a new methodology that is based on two reddening-free
observables: 1) a distance indicator called pseudomagnitude and 2) the 
differential surface brightness, a handy quasi-experimental
observable, that is independent of distance and specific to each star. 
This, together with our new database of measured angular diameters,
allows us to provide estimates of star diameters with 
statistical errors of $\sim\!1.1\%$, plus possible biases
of~$\sim\!2\%$.  It permits us to upgrade the JSDC catalog of
stellar diameters to about $453\,000$ stars, a tenfold improvement in
number of stars and diameter precision. 

The polynomial method developed in this work may be used with any photometric
system, selecting optical bands that best represent  the stellar
continuum. However, the exercise has severe limits because the DSB is
not smooth. The only way to reduce the biases and to go beyond $1\%$ error is to
measure its structure for all spectral types and luminosity
classes. This emphasises this importance to get precise photometry with
biases smaller than~$1\%$ and above all the importance of optical
interferometry to get precise and numerous angular diameters.
\begin{table}
\label{tab:coeffpol}
\begin{tabular}{lll|lll}
\hline
\T SP Type&$p$& $\sigma_p$ &  SP Type&$p$&$\sigma_p$ \B\\
\hline \hline
\T
O5& 0.2432 &    0.0443 &F6&0.5810 &   0.0025 \\
O6& 0.2534 &    0.0337 &F7&0.5822 &   0.0024 \\ 
O7& 0.2670 &    0.0255 &F8&0.5835 &   0.0023 \\
O8& 0.2830 &    0.0192 &F9&0.5850 &   0.0022 \\
O9& 0.3009 &    0.0147 &G0&0.5867 &   0.0021 \\
B0& 0.3201 &    0.0116 &G1&0.5888 &   0.0020 \\
B1& 0.3401 &   0.0096 &G2&0.5912 &   0.0020 \\
B2& 0.3603 &   0.0085 &G3&0.5939 &   0.0019 \\
B3& 0.3805 &   0.0077 &G4&0.5971 &   0.0019 \\
B4& 0.4003 &   0.0071 &G5&0.6006 &   0.0019 \\
B5& 0.4195 &   0.0067 &G6&0.6045 &   0.0019 \\
B6& 0.4378 &   0.0062 &G7&0.6088 &   0.0019 \\
B7& 0.4551 &   0.0059 &G8&0.6134 &   0.0020 \\
B8& 0.4712 &   0.0055 &G9&0.6183 &   0.0020 \\
B9& 0.4862 &   0.0053 &K0&0.6235 &   0.0021 \\
A0& 0.4998 &   0.0051 &K1&0.6289 &   0.0021 \\
A1& 0.5121 &   0.0050 &K2&0.6345 &   0.0022 \\
A2& 0.5231 &   0.0049 &K3&0.6402 &   0.0023 \\
A3& 0.5329 &   0.0048 &K4&0.6460 &   0.0025 \\
A4& 0.5414 &   0.0047 &K5&0.6518 &   0.0026 \\
A5& 0.5488 &   0.0046 &K6&0.6575 &   0.0028 \\
A6& 0.5551 &   0.0045 &K7&0.6632 &   0.0029 \\
A7& 0.5604 &   0.0043 &K8&0.6686 &   0.0030 \\
A8& 0.5648 &   0.0041 &K9&0.6739 &   0.0031 \\
A9& 0.5684 &   0.0039 &M0&0.6790 &   0.0029 \\
F0& 0.5714 &   0.0037 &M1&0.6838 &   0.0026 \\
F1& 0.5738 &   0.0034 &M2&0.6884 &   0.0023 \\
F2& 0.5757 &   0.0032 &M3&0.6928 &   0.0025 \\
F3& 0.5773 &   0.0030 &M4&0.6971 &   0.0039 \\
F4& 0.5786 &   0.0028 &M5&0.7012 &   0.0065 \\
F5& 0.5798 &   0.0026 &M6&0.7054 &   0.0104  \B \\ 
\hline
\end{tabular}
\caption{Differential surface brightness polynomial values $p$ and
  associated errors $\sigma_p$ for stars measured in the (V, Ks) photometric pair. The
  angular diameter (mas) and its statistical relative error may be
  obtained from Eq.~\ref{logtheta} and Eq.~\ref{error_theta},
  replacing X by Ks. 
}
\end{table}
%
%
\begin{table*}
\scriptsize
\begin{tabular}{
p{0.15\linewidth}|
p{0.15\linewidth}|
p{0.15\linewidth}|
p{0.15\linewidth}|
p{0.15\linewidth}|
p{0.15\linewidth}
}
\hline \hline
  Name \dotfill Ref(s) &
  Name \dotfill Ref(s) &
  Name \dotfill Ref(s) &
  Name \dotfill Ref(s) &
  Name \dotfill Ref(s) &
  Name \dotfill Ref(s) \B 
\\
\hline\T
GJ411 \dotfill  59&GJ412A \dotfill  75&GJ649 \dotfill  82&GJ687 \dotfill  75&HD100029 \dotfill  57, 63, 83&HD100920 \dotfill  19\\
HD1013 \dotfill  83&HD10144 \dotfill   9&HD101501 \dotfill  74&HD102212 \dotfill  43, 57, 63, 83&HD102328 \dotfill  72&HD102647 \dotfill   9, 66\\HD102870 \dotfill  74&HD103605 \dotfill  72&HD10380 \dotfill  36, 56&HD10476 \dotfill  77&HD104985 \dotfill  69&HD106574 \dotfill  72\\
HD106625 \dotfill   9&HD10697 \dotfill  69, 77&HD107383 \dotfill  82&HD10780 \dotfill  70, 74&HD108907 \dotfill  83&HD109358 \dotfill  74, 83\\
HD112300 \dotfill  63&HD113049 \dotfill  72&HD113226 \dotfill  56, 63, 83&HD113996 \dotfill  83&HD114710 \dotfill  74&HD114961 \dotfill  36, 43\\
HD115617 \dotfill  82&HD115659 \dotfill  80&HD117176 \dotfill  69&HD117675 \dotfill  83&HD118904 \dotfill  72&HD119149 \dotfill  23\\
HD11964 \dotfill  69, 77&HD11977 \dotfill  73&HD119850 \dotfill  75&HD120136 \dotfill  69&HD120477 \dotfill  56, 83&HD121130 \dotfill  83\\
HD121370 \dotfill  57, 63, 67&HD123139 \dotfill  61&HD123934 \dotfill  14, 16, 22&HD12479 \dotfill  14, 37, 50, 54&HD12533 \dotfill  32, 63&HD126660 \dotfill  74\\
HD127665 \dotfill  83&HD128167 \dotfill  74&HD12929 \dotfill  40, 46, 48, 56, 63&HD129712 \dotfill  83&HD130948 \dotfill  77&HD131156 \dotfill  74\\
HD131873 \dotfill  40, 63&HD13189 \dotfill  69&HD131977 \dotfill  65&HD132112 \dotfill  36, 43&HD1326 \dotfill  58, 59, 75&HD132813 \dotfill  63\\HD133124 \dotfill  83&HD133208 \dotfill  57, 63, 83&HD133774 \dotfill  23&HD135722 \dotfill  56, 63&HD136202 \dotfill  77&HD136726 \dotfill  72, 83\\
HD137443 \dotfill  72&HD137759 \dotfill  83&HD138265 \dotfill  72&HD139357 \dotfill  72&HD139663 \dotfill  12&HD140538 \dotfill  77\\
HD140573 \dotfill  57, 63, 83&HD141795 \dotfill  74&HD142804 \dotfill  36&HD142860 \dotfill  74&HD143107 \dotfill  83&HD143761 \dotfill  69, 82\\
HD144690 \dotfill  50&HD145675 \dotfill  69&HD146051 \dotfill  63, 78&HD146233 \dotfill  74&HD148387 \dotfill  57, 63, 83&HD148478 \dotfill  47, 63\\
HD149661 \dotfill  75&HD150383 \dotfill  49&HD150680 \dotfill  57, 63&HD150798 \dotfill  78&HD150997 \dotfill  56, 63, 83&HD1522 \dotfill  83\\
HD152786 \dotfill  78&HD154345 \dotfill  71&HD156283 \dotfill  46, 56, 63&HD157214 \dotfill  77&HD157681 \dotfill  72&HD158633 \dotfill  77\\
HD159181 \dotfill  63&HD159561 \dotfill   9&HD160290 \dotfill  72&HD161096 \dotfill  83&HD16141 \dotfill  71&HD16160 \dotfill  58, 59, 75\\
HD161797 \dotfill  63&HD162003 \dotfill  74&HD163770 \dotfill  63&HD163917 \dotfill  83&HD164058 \dotfill  40, 41, 44, 46, 63&HD164259 \dotfill  74\\
HD167042 \dotfill  72&HD16765 \dotfill  77&HD168151 \dotfill  77&HD16895 \dotfill  74&HD168988 \dotfill  36&HD169022 \dotfill   9\\
HD170693 \dotfill  72, 83&HD172167 \dotfill   1,  9, 63&HD172816 \dotfill  17, 23, 29, 34, 36, 37, 55&HD17361 \dotfill  56&HD173667 \dotfill  74&HD17506 \dotfill  63\\
HD175588 \dotfill  51, 63&HD175726 \dotfill  76&HD175775 \dotfill  18&HD175823 \dotfill  72&HD175865 \dotfill  51, 63&HD176124 \dotfill  36\\
HD176408 \dotfill  72&HD176411 \dotfill  56&HD176437 \dotfill  79, 81&HD176524 \dotfill  83&HD176678 \dotfill  83&HD17709 \dotfill  63\\
HD177153 \dotfill  76&HD177724 \dotfill  74&HD177756 \dotfill  81&HD177830 \dotfill  69&HD180540 \dotfill  34, 55&HD180610 \dotfill  83\\
HD180711 \dotfill  40, 57, 63&HD180809 \dotfill  46, 56&HD181276 \dotfill  56, 83&HD181420 \dotfill  76&HD18191 \dotfill  16, 35&HD182572 \dotfill  74\\
HD183439 \dotfill  46, 63, 83&HD184171 \dotfill  81&HD185144 \dotfill  70, 74&HD185395 \dotfill  74&HD186408 \dotfill  77&HD186427 \dotfill  69, 77\\
HD186791 \dotfill  56, 63&HD186815 \dotfill  72&HD186882 \dotfill  81&HD187082 \dotfill  23&HD187637 \dotfill  76&HD188512 \dotfill  56\\
HD190228 \dotfill  69&HD190360 \dotfill  69&HD190406 \dotfill  76&HD19058 \dotfill  44, 51, 63&HD192781 \dotfill  72&HD19373 \dotfill  74\\
HD193924 \dotfill   1,  9&HD194093 \dotfill  56, 63&HD195564 \dotfill  77&HD195820 \dotfill  72&HD196777 \dotfill   4, 30, 36&HD197345 \dotfill  32, 63\\
HD19787 \dotfill  56&HD197989 \dotfill  48, 63&HD198149 \dotfill  56&HD199305 \dotfill  75&HD199665 \dotfill  71&HD19994 \dotfill  69\\
HD200205 \dotfill  72&HD200905 \dotfill  56, 63&HD201092 \dotfill  68&HD202109 \dotfill  63&HD202850 \dotfill  79&HD203504 \dotfill  83\\
HD204724 \dotfill  63&HD205435 \dotfill  56&HD20630 \dotfill  74&HD206778 \dotfill  57, 63&HD206860 \dotfill  77&HD206952 \dotfill  83\\
HD207005 \dotfill  12&HD20902 \dotfill  32, 56, 63&HD209100 \dotfill  65&HD209750 \dotfill  56, 63&HD209950 \dotfill  29&HD209952 \dotfill   1,  9\\
HD210027 \dotfill  76&HD21019 \dotfill  77&HD210418 \dotfill  74&HD210702 \dotfill  71, 82&HD210745 \dotfill  63&HD212496 \dotfill  56, 83\\
HD213306 \dotfill  53, 56&HD213558 \dotfill  74&HD214868 \dotfill  56, 72&HD214923 \dotfill  81&HD215648 \dotfill  74&HD215665 \dotfill  56, 63\\
HD216032 \dotfill  43&HD216131 \dotfill  56, 63&HD216386 \dotfill   2, 63&HD216956 \dotfill   1,  9, 66&HD217014 \dotfill  69, 77&HD217906 \dotfill  44, 46, 48, 51, 52, 63, 80\\
HD218329 \dotfill  83&HD218356 \dotfill  63&HD218396 \dotfill  76&HD219080 \dotfill  79&HD219134 \dotfill  75&HD219576 \dotfill  36\\
HD219615 \dotfill  83&HD219623 \dotfill  77&HD221115 \dotfill  56&HD221345 \dotfill  71&HD222368 \dotfill  74&HD222603 \dotfill  77\\
HD224062 \dotfill  19, 37&HD22484 \dotfill  74&HD224935 \dotfill  64&HD23249 \dotfill  67&HD23319 \dotfill  73&HD23596 \dotfill  69\\
HD24398 \dotfill  81&HD24512 \dotfill  78&HD25025 \dotfill  63&HD25604 \dotfill  56&HD25705 \dotfill  78&HD27256 \dotfill  73\\
HD285968 \dotfill  82&HD29139 \dotfill   8, 24, 25, 26, 27, 28, 33, 38, 41, 42, 44, 48, 51, 63&HD30959 \dotfill  78&HD31398 \dotfill  63&HD31767 \dotfill  63&HD31964 \dotfill  63\\
HD32518 \dotfill  72&HD32630 \dotfill  79&HD33564 \dotfill  82&HD3360 \dotfill  79&HD33793 \dotfill  59&HD34085 \dotfill   1,  9\\
HD34411 \dotfill  74&HD3546 \dotfill  56&HD35468 \dotfill   1,  9, 81&HD3627 \dotfill  48, 57, 63&HD36389 \dotfill  31, 36, 37, 39, 78&HD36395 \dotfill  59, 75\\
HD3651 \dotfill  69&HD36673 \dotfill  56&HD36848 \dotfill  73&HD3712 \dotfill  32, 46, 48, 56, 63&HD37128 \dotfill   1,  9&HD38529 \dotfill  69\\
HD38858 \dotfill  77&HD38944 \dotfill  83&HD39983 \dotfill  43, 54&HD40239 \dotfill  51, 63&HD4128 \dotfill  78&HD42995 \dotfill  60\\
HD432 \dotfill  56&HD44478 \dotfill   6,  7, 10, 11, 21, 35, 41, 44, 51, 63&HD45348 \dotfill   1,  9, 78&HD45410 \dotfill  71&HD4628 \dotfill  75&HD4656 \dotfill  50\\
HD48329 \dotfill  13, 43, 45, 50, 56, 63&HD48915 \dotfill   1,  9, 62, 63&HD49933 \dotfill  76&HD49968 \dotfill  30&HD5015 \dotfill  74&HD52089 \dotfill   1,  9\\
HD5395 \dotfill  56&HD5448 \dotfill  79&HD54605 \dotfill   9&HD54719 \dotfill  83&HD56537 \dotfill  74&HD57423 \dotfill  35\\
HD5820 \dotfill  23&HD58350 \dotfill   9&HD58946 \dotfill  74&HD59686 \dotfill  69&HD60294 \dotfill  72&HD6210 \dotfill  77\\
HD62345 \dotfill  83&HD66141 \dotfill  83&HD66811 \dotfill   3,  9&HD6860 \dotfill  32, 41, 44, 46, 48, 51, 63&HD69267 \dotfill  56, 63&HD69897 \dotfill  77\\
HD70272 \dotfill  46&HD7087 \dotfill  56&HD73108 \dotfill  72&HD74442 \dotfill  36&HD76294 \dotfill  57, 83&HD76827 \dotfill  63\\
HD79211 \dotfill  75&HD7924 \dotfill  82&HD80007 \dotfill   9&HD8019 \dotfill  23&HD80493 \dotfill  46, 57, 63&HD81797 \dotfill  63, 78\\
HD81937 \dotfill  74&HD82308 \dotfill  83&HD82885 \dotfill  74&HD83618 \dotfill  83&HD84194 \dotfill  83&HD84441 \dotfill  57, 63\\
HD8512 \dotfill  83&HD85503 \dotfill  83&HD86663 \dotfill  18, 23, 56&HD86728 \dotfill  74&HD87837 \dotfill  13, 15, 21, 29, 56&HD87901 \dotfill   1,  9\\
HD88230 \dotfill  58, 59, 75&HD89449 \dotfill  76, 79&HD89758 \dotfill  57, 63&HD90839 \dotfill  74&HD91232 \dotfill  43&HD9138 \dotfill  45\\
HD9408 \dotfill  56&HD95418 \dotfill  74&HD95608 \dotfill  79&HD95735 \dotfill  58, 58, 75, 75&HD96833 \dotfill  57, 63, 83&HD97603 \dotfill  74\\HD97633 \dotfill  79&HD9826 \dotfill  69&HD98262 \dotfill  63&HD9927 \dotfill  46, 56&HD99998 \dotfill   5, 20, 56&HR4518 \dotfill  56\\
HR9045 \dotfill  56\B\\
\hline
\end{tabular}
\caption{\label{starlist}List of stars and associated references of
  angular diameter measurements that have been retained to calibrate
  the DSB in this work.} 
\tablebib{
(1)\,\citet{ 1967MNRAS.137..393H}; (2)\,\citet{ 1970ApJ...160L.181N}; (3)\,\citet{ 1970MNRAS.150...45D}; (4)\,\citet{ 1973AJ.....78..199D}; (5)\,\citet{ 1974AJ.....79..483D}; (6)\,\citet{ 1974AJ.....79.1076W}; (7)\,\citet{ 1974AJ.....79.1079R}; (8)\,\citet{ 1974ApJ...187..131C}; (9)\,\citet{ 1974MNRAS.167..121H}; (10)\,\citet{ 1975AJ.....80...45D}; (11)\,\citet{ 1975ApJ...198..127N}; (12)\,\citet{ 1975MNRAS.170..229H}; (13)\,\citet{ 1976A&A....47..457D}; (14)\,\citet{ 1976AJ.....81..650A}; (15)\,\citet{ 1976MNRAS.175p..57G}; (16)\,\citet{ 1977AJ.....82..414R}; (17)\,\citet{ 1977AJ.....82..631A}; (18)\,\citet{ 1977PASP...89...95V}; (19)\,\citet{ 1978AJ.....83.1100A}; (20)\,\citet{ 1978AJ.....83.1639W}; (21)\,\citet{ 1978AN....299..243B}; (22)\,\citet{ 1978IAUS...80..447W}; (23)\,\citet{ 1979AJ.....84..247R}; (24)\,\citet{ 1979AJ.....84..872W}; (25)\,\citet{ 1979ApJ...228L.111B}; (26)\,\citet{ 1979MNRAS.187..753B}; (27)\,\citet{ 1980AJ.....85...47P}; (28)\,\citet{ 1980AJ.....85.1262E}; (29)\,\citet{ 1980AJ.....85.1496R}; (30)\,\citet{ 1980AJ.....85.1505B}; (31)\,\citet{ 1980ApJ...242..646W}; (32)\,\citet{ 1981A&A...103...28B}; (33)\,\citet{ 1981AJ.....86..906R}; (34)\,\citet{ 1981AJ.....86.1277E}; (35)\,\citet{ 1981AJ.....86.1404B}; (36)\,\citet{ 1982AJ.....87..808R}; (37)\,\citet{ 1982AJ.....87..818B}; (38)\,\citet{ 1982AJ.....87.1044R}; (39)\,\citet{ 1982ApJ...254..670W}; (40)\,\citet{ 1983A&A...120..263F}; (41)\,\citet{ 1983ApJ...268..309D}; (42)\,\citet{ 1984AJ.....89..424W}; (43)\,\citet{ 1986AJ.....91..961S}; (44)\,\citet{ 1987A&A...188..114D}; (45)\,\citet{ 1987AJ.....94..201S}; (46)\,\citet{ 1989ApJ...340.1103H}; (47)\,\citet{ 1990A&A...230..355R}; (48)\,\citet{ 1991AJ....101.2207M}; (49)\,\citet{ 1992A&A...254..149R}; (50)\,\citet{ 1992A&A...265..535R}; (51)\,\citet{ 1993ApJ...406..215Q}; (52)\,\citet{ 1995AJ....109..378D}; (53)\,\citet{ 1997A&A...317..789M}; (54)\,\citet{ 1997MNRAS.287..681R}; (55)\,\citet{ 1998A&A...330..578R}; (56)\,\citet{ 1999AJ....118.3032N}; (57)\,\citet{ 2001AJ....122.2707N}; (58)\,\citet{ 2001ApJ...551L..81L}; (59)\,\citet{ 2003A&A...397L...5S}; (60)\,\citet{ 2003A&A...399..275R}; (61)\,\citet{ 2003A&A...404.1087K}; (62)\,\citet{ 2003A&A...408..681K}; (63)\,\citet{ 2003AJ....126.2502M}; (64)\,\citet{ 2004A&A...419..285F}; (65)\,\citet{ 2004A&A...426..297K}; (66)\,\citet{ 2004A&A...426..601D}; (67)\,\citet{ 2005A&A...436..253T}; (68)\,\citet{ 2008A&A...488..667K}; (69)\,\citet{ 2008ApJ...680..728B}; (70)\,\citet{ 2008ApJ...683..424B}; (71)\,\citet{ 2009ApJ...701..154B}; (72)\,\citet{ 2010ApJ...710.1365B}; (73)\,\citet{ 2012A&A...539A..58C}; (74)\,\citet{ 2012ApJ...746..101B}; (75)\,\citet{ 2012ApJ...757..112B}; (76)\,\citet{ 2013.TB.private}; (77)\,\citet{ 2013ApJ...771...40B}; (78)\,\citet{ 2013MNRAS.434..437C}; (79)\,\citet{ 2013MNRAS.434.1321M}; (80)\,\citet{ 2014A&A...566A..88A}; (81)\,\citet{ 2014A&A...570A.104C}; (82)\,\citet{ 2014MNRAS.438.2413V}; (83)\,\citet{ 2014SPIE.9146E..0WB}}
\end{table*}
\begin{acknowledgements} 
  This research has made use of NASA's Astrophysics Data System.  This
  research has made use of the SIMBAD database
  \citep{2000A&AS..143....9W}, and of the VizieR catalog access
  tool \citep{2000A&AS..143...23O}, CDS, Strasbourg, France. The
  TOPCAT tool\footnote{available at
    \url{http://www.starlink.ac.uk/topcat/}}\citep{2005ASPC..347...29T}
  was pivotal in the analysis and filtering of our databases. This publication makes use of data products from the Two Micron All Sky Survey, which is a joint project of the University of Massachusetts and the Infrared Processing and Analysis Center/California Institute of Technology, funded by the National Aeronautics and Space Administration and the National Science Foundation.
\end{acknowledgements}
\nocite{*}
\bibliographystyle{aa} 
\bibliography{chelli_paper1_v6} 
\begin{appendix}
\section{Linear least squares fit}
\label{app:linearsqfit1}
The data consist of a set of $N_S$ stars that are characterized
by their measured diameter $\theta$ (corrected for limb-darkening) and
$N_B$ observed magnitudes. Each star provides $N_B-1$ 
linearly independent pseudomagnitude, which combined with the diameter
$\theta$, give $N_B-1$ correlated measurements (differential surface
brightness) per star. Each measurement is fitted as a function of 
the stellar spectral type number (from 0 to 69 for O$0$ to M$9$) with
a polynomial of degree $m$. The problem is then to 
evaluate $(m+1) \times (N_B-1)$ polynomial coefficients, which we do
via a simple linear least squares fit, described below.
\subsection{Polynomials calculation}
\label{app:linearsqfit}
We align the $(N_B-1) \times N_S$ measurements in a vector ${\mathbf
  M}$, and we define the transition matrix ${\mathbf T}$ of dimensions
$(N_B-1).N_S \times (m+1).(N_B-1)$ as the derivative of ${\mathbf
  M}$ with respect to the unknowns. For
simplicity, we assume that the $N_S$ stars are distinct, implying no correlations
between the measurements of different stars. This allows us  to define
$N_S$ independent covariance matrices $\mathbf {C}_i,
i=0,1,...,N_S-1$,of dimension $(N_B-1) \times (N_B-1)$. Given the
photometric pairs that were selected in 
this work (V, J),(V, H),(V, Ks), the generic expression of the
covariance matrix is  
\begin{eqnarray}
& \mathbf{C} = \frac{\sigma_\theta^2}{\theta^2 \ln(10)^2} +
0.04 \times ,\\
& \left[
\begin{array}{ccc}
\frac{c_J^2\sigma_V^2+c_V^2\sigma_J^2}{(c_V-c_J)^2}
& \frac{c_Jc_H\sigma_V^2}{(c_V-c_J)\times(c_V-c_H)} & \frac{c_Jc_{Ks}\sigma_V^2}{(c_V-c_J)\times(c_V-c_{Ks})}\\
\frac{c_J\,c_H\sigma_V^2}{(c_V-c_J)\times(c_V-c_H)}
& \frac{c_H^2\sigma_V^2+c_V^2\sigma_H^2}{(c_V-c_H)^2} & \frac{c_Hc_{Ks}\sigma_V^2}{(c_V-c_H)\times(c_V-c_{Ks})}\\
\frac{c_Jc_{Ks}\sigma_V^2}{(c_V-c_J)\times(c_V-c_{Ks})} & \frac{c_Hc_{Ks}\sigma_V^2}{(c_V-c_H)\times(c_V-c_{Ks})} &
\frac{c_{Ks}^2\sigma_V^2+c_V^2\sigma_{Ks}^2}{(c_V-c_{Ks})^2} \\
\nonumber
\end{array}
\right]
\end{eqnarray} 
where $\theta$ is the measured diameter and $\sigma_\theta$ its error,
$(\sigma_V,\sigma_J,\sigma_H,\sigma_{Ks})$,  
and  $(c_V,c_J,c_H,c_{Ks})$ are the magnitude errors and the
interstellar extinction coefficients in the corresponding bands.
Next we place the inverse of the covariance matrices $\{\mathbf{C}_i^{-1}\}$
along the diagonal of a matrix $\mathbf{D}$ of dimensions $(N_B-1).N_S \times
(N_B-1).N_S$. The solution for the polynomial coefficients is
contained in a vector $\mathbf{A}$ of dimensions $(m+1) \times (N_B-1)$, given by
\begin{equation}
\label{eq:polynom_coefficients}
\mathbf{A}=\Big [ ^t\mathbf{T} \times \mathbf{D} \times \mathbf{T} \Big]^{-1} \times 
^t\mathbf{T} \times \mathbf{D} \times \mathbf{V}
.\end{equation}
The covariance matrix $\mathbf{C}_a$ of the solution $\mathbf{A}$ is
\begin{equation}
\mathbf{C}_a=\Big [ ^t\mathbf{T} \times \mathbf{D} \times \mathbf{T} \Big ]^{-1}
.\end{equation}
The reconstructed measurement vector writes: $\mathbf{M}_r =
  \mathbf{T} \times \mathbf{A}$, and the reduced $\chi_p^2$ of the
  fitting 
process is
\begin{equation}
\label{eq:reduced_chisquare}
\chi_p^2= \frac {^t(\mathbf{M}-\mathbf{M}_r)\times \mathbf{D}\times (\mathbf{
     M}-\mathbf{M}_r)} {N_S\times (N_B-1)}
.\end{equation}
\subsection{Diameter calculation}
\label{app:diamcalc}
For a given star with a spectral type number $n_s$ and an associated
pseudomagnitude vector $\mathbf{P}$ 
of dimension $N_B-1$, the reconstructed {\it iest} $({\it i}=0,...N_B-1)$
diameter $\theta_i$ is given by
\begin{equation}
\log(\theta_i) = \sum_{k=0}^{k=m} \mathbf{A}\big [k+i\times (m+1) \big] \times
n_s^k-0.2\times \mathbf{P}(i)
.\end{equation}
The covariance matrix $\mathbf{C}_d$ between log diameter estimates writes%
\begin{eqnarray}
\mathbf{C}_d\big[\log(\theta_i),\log(\theta_j)\big] = 0.04 \,\mathrm
{cov}\big [\mathbf{P}(i),\mathbf{P}(j) \big ] +
.\\ \nonumber 
\sum_{k=0}^{k=m}\sum_{l=0}^{l=m} \mathbf{C}_a\big[k+i\times (m+1),l+j\times
(m+1) \big]\times n_s^{k+l}
\end{eqnarray}
Let us range the $N_B-1$ log diameter estimates within a vector
${\mathbf R}$. The mean log diameter $\overline {\log (\theta)}$ and the
associated error are
\begin{equation}
{\overline {\log (\theta)}} = \frac {\sum \mathbf{C}_d^{-1}
  \times \mathbf{R}}{\sum \mathbf{C}_d^{-1}}
\end{equation}
\begin{equation}
\sigma \big[\overline {\log (\theta)}\big] = \Big [ \sum \mathbf{C}_d^{-1} \Big]^{-0.5}
,\end{equation}
where $\sum$ stands for the sum of all the matrix elements. The mean 
diameter $\overline{\theta}$ and its error are computed as follows:
\begin{equation}
\overline{\theta} = 10^{\overline {\log (\theta)}}
\end{equation}
\begin{equation}
\sigma(\overline{\theta})= \ln (10) \, \sigma \big[\overline {\log (\theta)}\big]\times \overline{\theta}  
.\end{equation}
At last, we define the chi-square $\chi^2_\theta$ associated with the reconstructed log
diameter (from the database) by
\begin{equation}
\chi^2_\theta = \frac{^t\mathbf{B}\times[\mathbf{C}_d+\sigma_\theta^2]^{-1}\times \mathbf{B}}{N_B-1}
,\end{equation}
where $\mathbf{B}$ is the vector of the difference between the log
of the diameter estimates ($\mathbf{R}$) and that of the measured
diameter, and $\sigma_\theta$ is the error of the measured log diameter. In the case of a catalog of stars with no measured diameter, we can
define an internal $\chi_\theta^2$, replacing the measured diameter and
its error with the mean computed diameter and its error.  
\end{appendix}
\end{document}